\documentclass[dvipsnames,format=sigconf,anonymous=false,review=false]{acmart} %% anonymous 
\AtBeginDocument{%
  \providecommand\BibTeX{{%
    \normalfont B\kern-0.5em{\scshape i\kern-0.25em b}\kern-0.8em\TeX}}}

\setcopyright{acmcopyright}
\copyrightyear{2023}
\acmYear{2023}
\acmConference[GECCO '123]{GECCO '23: The Genetic and Evolutionary Computation Conference}{July 15-19, 2023}{Lisbon, Portugal}
\acmISBN{978-1-4503-XXXX-X/18/06}

\usepackage{hyperref}
\usepackage{subfigure}
\usepackage{float}

\usepackage{algorithm} 
\usepackage{algpseudocode}

\usepackage[export]{adjustbox}
\usepackage{tikz}
\usetikzlibrary{positioning, quotes}

\usepackage{xcolor}
\definecolor{viridis0}{HTML}{fde725}
\definecolor{viridis1}{HTML}{5ec962}
\definecolor{viridis2}{HTML}{21918c}
\definecolor{viridis3}{HTML}{3b528b}
\definecolor{viridis4}{HTML}{440154}

\copyrightyear{2023} 
\acmYear{2023} 
\setcopyright{rightsretained} 
\acmConference[GECCO '23 Companion]{Genetic and Evolutionary Computation Conference Companion}{July 15--19, 2023}{Lisbon, Portugal}
\acmBooktitle{Genetic and Evolutionary Computation Conference Companion (GECCO '23 Companion), July 15--19, 2023, Lisbon, Portugal}\acmDOI{10.1145/3583133.3590744}
\acmISBN{979-8-4007-0120-7/23/07}

\begin{document}

%%
%% The "title" command has an optional parameter,
%% allowing the author to define a "short title" to be used in page headers.
\title{Coping with seasons: evolutionary dynamics of gene networks in a changing environment}

%%
%% The "author" command and its associated commands are used to define
%% the authors and their affiliations.
%% Of note is the shared affiliation of the first two authors, and the
%% "authornote" and "authornotemark" commands
%% used to denote shared contribution to the research.

\author{Csenge Petak}
\authornote{Both authors contributed equally to this research.}
\email{cpetak@uvm.edu}
\affiliation{
  %\institution{Pespeni Lab}
  \institution{University of Vermont}
  \city{Burlington}
  \state{VT}
  \country{USA}
}

\author{Lapo Frati}
\authornotemark[1]
\email{lfrati@uvm.edu}
%\orcid{1234-5678-9012}
\affiliation{%
  %\institution{Neurobotics Lab}
  \institution{University of Vermont}
  \streetaddress{P.O. Box 1212}
  \city{Burlington}
  \state{VT}
  \country{USA}
}

\author{Melissa H. Pespeni}
\email{mpespeni@uvm.edu}
\affiliation{
  %\institution{Pespeni Lab}
  \institution{University of Vermont}
  \city{Burlington}
  \state{VT}
  \country{USA}
}

\author{Nick Cheney}
\email{ncheney@uvm.edu}
\affiliation{
  %\institution{Neurobotics Lab}
  \institution{University of Vermont}
  \city{Burlington}
  \state{VT}
  \country{USA}
}

%%
%% By default, the full list of authors will be used in the page
%% headers. Often, this list is too long, and will overlap
%% other information printed in the page headers. This command allows
%% the author to define a more concise list
%% of authors' names for this purpose.
\renewcommand{\shortauthors}{Petak, Frati et al.}

%%
%% The abstract is a short summary of the work to be presented in the
%% article.
\begin{abstract}
In environments that vary frequently and unpredictably, bet-hedgers can overtake the population. Diversifying bet-hedgers have a diverse set of offspring so that, no matter the conditions they find themselves in, at least some offspring will have high fitness. In contrast, conservative bet-hedgers have a set of offspring that all have an in-between phenotype compared to the specialists. Here, we use an evolutionary algorithm of gene regulatory networks to de novo evolve the two strategies and investigate their relative success in different parameter settings. We found that diversifying bet-hedgers almost always evolved first, but then eventually got outcompeted by conservative bet-hedgers. We argue that even though similar selection pressures apply to the two bet-hedger strategies, conservative bet-hedgers could win due to the robustness of their evolved networks, in contrast to the sensitive networks of the diversifying bet-hedgers. These results reveal an unexplored aspect of the evolution of bet-hedging that could shed more light on the principles of biological adaptation in variable environmental conditions.
\end{abstract}

%%
%% The code below is generated by the tool at http://dl.acm.org/ccs.cfm.
%% Please copy and paste the code instead of the example below.
%%

%\ccsdesc[100]{Networks~Network reliability}

%%
%% Keywords. The author(s) should pick words that accurately describe
%% the work being presented. Separate the keywords with commas.

%%
%% This command processes the author and affiliation and title
%% information and builds the first part of the formatted document.

\maketitle

\section{Introduction}

No environment ever stays the same. What do you do when you can’t predict what happens next? You hedge your bets to maximize your long-term survival. As it turns out, this is exactly what populations in nature evolve to do as well. 
Environmental variability comes in all shapes and sizes. Depending on whether the environment changes during the lifetime of an individual or once every thousand generations, and whether the changes are predictable and the cues are accurate, populations can adapt to these changes in the environment using different strategies \cite{Simons2011-id}. If the environmental change is infrequent the population likely goes through a process called adaptive tracking, during which the population continuously adapts. Specialists (i.e., only fit in one of
the reoccurring environments) for one environment get outcompeted by specialists for the other environment every time the environment changes. On the other hand, if the change is frequent and there is a reliable environmental cue, phenotypic plasticity can evolve, which is the expression of an alternative phenotype in response to an environmental cue. However, in many cases, there is no reliable cue to signal the environmental change to the individual. In these cases, a strategy call bet-hedging can evolve \cite{Botero2015-ol}.

A biological bet-hedging strategy is one that results in a decreased variance among the fitness of the offspring across all possible environmental conditions compared to the specialists. There are two main ways to achieve this: 1) \textit{diversifying bet-hedgers} (BHs) increase phenotypic diversity resulting in different phenotypes among the offspring that are fit in different environmental conditions; 2) \textit{conservative} BHs adopt a generalist phenotype that is somewhat fit in all environments \cite{seger1987bet}. Thus, when the environment is stable, bet-hedgers are selected against. However, when the environment switches, BH strategies have an advantage over the specialist. Thus, over many environmental switches, BHs can raise in frequency in the population \cite{Gillespie1974-hu, philippi1989hedging}.

\begin{figure*}
    \centering
    \includegraphics[width=\linewidth]{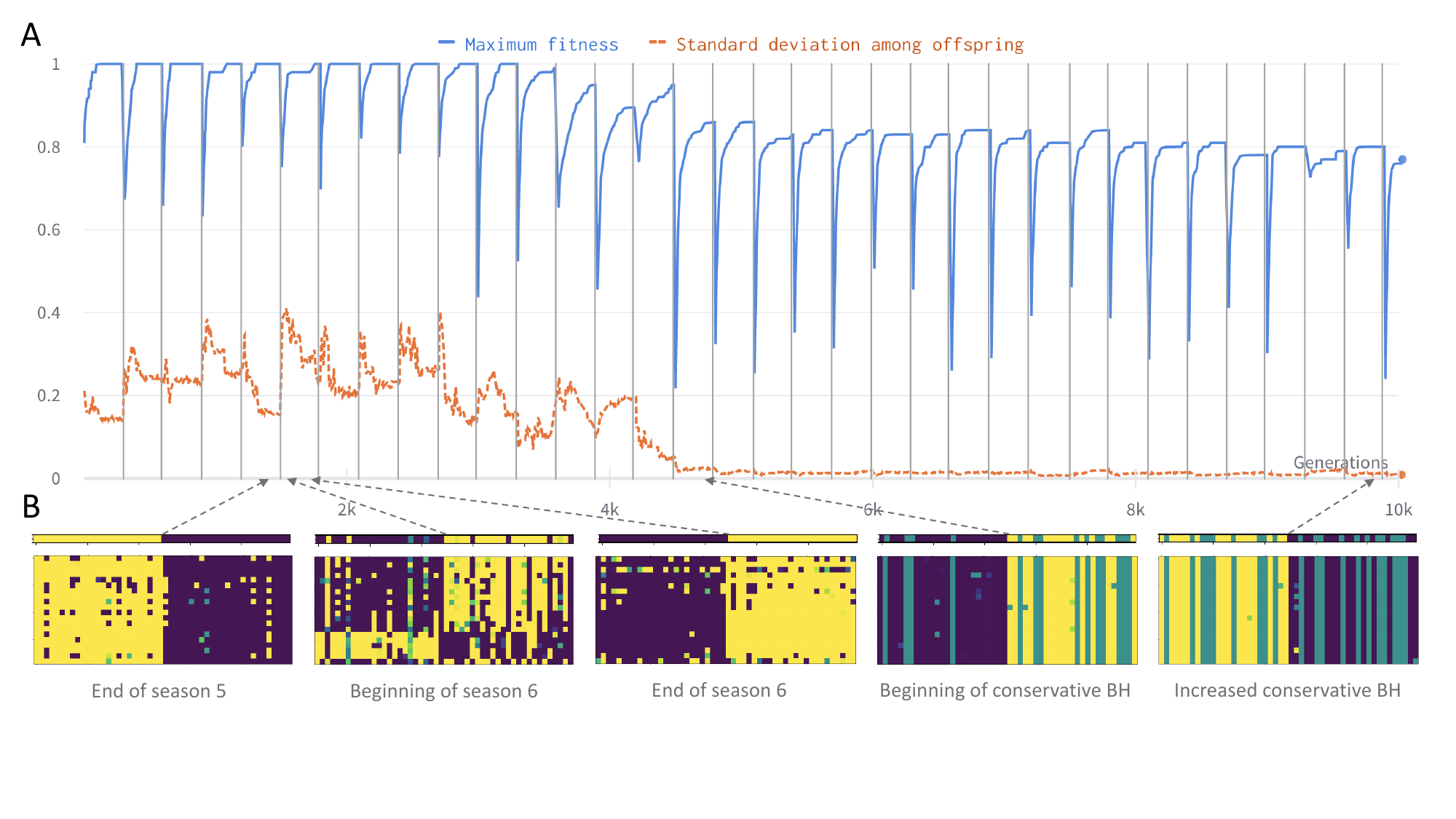}
    \vspace*{-20mm}
    \caption{A) Maximum fitness (blue) and average standard deviation among offspring of the same parent (orange) decreased over time. Grey vertical line: target switch. B) Phenotypes of the highest fitness parents along with 20 of their offspring. Rows: offspring, columns: genes, colors: expression level. Gene is off = purple, on = yellow, half expressed = turquoise. Example phenotypes in order of appearance: specialist, diversifying BH, specialist, specialist, conservative BH. }
    \label{max_std}
\end{figure*}

Examples of bet-hedging strategies in nature span across 16 phyla over 100 studies \cite{Simons2011-id}. One of the most common examples is the timing of insect diapause. Many species of insects grow exponentially during a growing season but produce an overwintering  alternative phenotype (cold resistant with arrested growth and reproduction) occasionally as to ensure survival when the environment suddenly turns cold \cite{Joschinski2020-ye}. Other examples include galactose metabolism in yeast \cite{acar2005enhancement}, persistence to antibiotics \cite{balaban2004bacterial} and even cancer cells \cite{gupta2011stochastic}. There has been much theoretical work using mathematical and agent-based models to understand the conditions and scenarios in which the different kinds of bet-hedging strategies could evolve. The general consensus among these studies is that while the two bet-hedging strategies are fundamentally similar in how and when they evolve \cite{Starrfelt2012-hl, Liu2019-jj, Haaland2020-mt}, higher frequency of environmental change favors the conservative \cite{Botero2015-ol, Mayer2017-lm}, and stronger selection pressure favors the diversifying BH strategy \cite{Tufto2015-lp, bull1987evolution}. However, these models didn't include a complex genotype-to-phenotype mapping function, and thus the strategies weren't evolved from scratch. Instead, the probability of producing an alternative phenotype like a diversifying BH was part of the genotype as an explicit evolvable variable \cite{Botero2015-ol}. 
 
The phenotypes of biological organisms are determined from their genotypes through a complex nonlinear mapping. Models of gene regulatory networks (GRNs, where nodes represent genes and edges represent activating or repressing directional regulatory interactions) are commonly used as conceptual proxies to genotype-to-phenotype mapping functions. Since the structure of GRNs are evolvable and shape the kind of phenotypic variation that is available for natural selection, they are often used in studies investigating the evolution of robustness and evolvability \cite{crombach2008evolution, Roh2013-xx,draghi2009evolutionary}. 

In this study, instead of using traditional mathematical models, we used an evolutionary algorithm to model the evolution of GRNs to investigate the emergence and success of diversifying and conservative BH strategies under different conditions. This approach allowed us to find these strategies without biasing or limiting our model; the strategies evolved without an explicit incentive through the evolution of different network structures, which led to some unexpected results.

\section{Methods}

\subsection{Genotypes and phenotypes}
In most computational models of GRNs, regulatory interactions between genes are simulated by an adjacency matrix $W$ of size $N\times N$, representing a weighted, directed graph. Then, the expression level of the genes making up the phenotype are calculated through the iterative multiplication of $W$ by a vector of gene expression levels $\vec{p}$ with a non-linear transformation. In our model, the initial vector $\vec{p}$ (representing gene products coming from the parent, i.e., maternal factors) was a one-hot vector of length $N = 50$ in all experiments. In order to generate a phenotype for each individual, this fixed input vector was iteratively multiplied a 100 times by their individual matrices as such:

\begin{align*}
\vec{p}_{t+1}&=\sigma\left(W\vec{p_{t}}\right) \\
%g{i}(t):=\sum_{j=1}^{N} w j} \vec{P}_{j}(t) \\
\sigma\left(x\right)&=\frac{1}{1+e^{-10x}}
\end{align*}

\noindent Where $W\vec{p}$ describes the ``strength" of interaction between genes and $\sigma$ is a sigmoid function. The value of $\vec{p}$ is bounded between 0 and 1. The individuals' phenotype was the value of gene expression levels $\vec{p}$ after this iterative process. 

\subsection{The evolutionary algorithm}

At the beginning of each experiment, a population of a 1000 haploid, asexual individuals was generated along with the two target vectors $\vec{A}$ and $\vec{B}$. $\vec{A}$ was a series of $N/2$ 1s followed by $N/2$ 0s, and $\vec{B}$ was 1 - $\vec{A}$, see \textbf{Fig \ref{max_std}B} first and third example phenotypes. Each experiment started with season A, during which the fitness of the individuals was calculated based on their distance from $\vec{A}$:

\begin{align*}
    f_A(\vec{p})= 1 - \dfrac{\sum\limits^{N} \lvert \vec{p}_{i} - \vec{A}_{i} \rvert }{N}
\end{align*}
Every G generations (season length) the target changed from $\vec{A}$ to $\vec{B}$ or from $\vec{B}$ to $\vec{A}$. After each individual's phenotype and fitness was calculated, they were sorted based on fitness and the top $\mu$ individuals were selected to survive to the next generation and generate $(popsize/\mu)-1$ offspring to keep a constant population size ($\mu$ + $\lambda$ Evolution Strategy). Offspring were mutated at $N*m$ positions by adding a random value drawn from a normal distribution ${\sim}\mathcal{N}(0,0.5)$. 

\subsection{Experiment}

Experiments were run for 75 environmental switches for 6 different season lengths (G): 20, 50, 100, 300, 400, 500, and 3 mutation rates ($m$) and truncation sizes ($\mu$/pop size): 0.05, 0.1, 0.2 for season lengths 50, 300 and 500. Each combination of parameters was repeated 10 times.

In order to calculate the mutational robustness of an evolved diversifying and conservative BH network, we mutated and evaluated the networks cumulatively 20 times. At each step, we quantified how much of a specialist, diversifying, and conservative BH they were based on the following coefficients: the diversifying BH coefficient was the standard deviation of offspring fitnesses given one of the targets, the conservative BH coefficient was the proportion of genes that are half expressed, and the specialist coefficient was the maximum between the average fitness of the offspring calculated for each of the two environments.

Code is available at: \url{https://github.com/Cpetak/coping_with_seasons_GRN}

\section{Results}

Populations evolved to have a lower decrease in average fitness upon a switch in the environment in all of our experiments. The maximum fitness the population reached by the end of a season also decreased over time, \textbf{Fig \ref{max_std}A}. This was due to the slow but steady incorporation of in-between 0.5 values to the phenotype, meaning that instead of genes being on or off, more and more genes were half expressed in the individuals over the generations. Therefore, by the end of most experiments, \textbf{conservative BHs} took over the population, \textbf{Fig \ref{max_std}B} right most example phenotype.

In the majority of the runs across different settings of the parameters, we also observed the raise and eventual fall of the \textbf{diversifying BH}. Our results showed that during approximately the first third of the simulations, for a short period of time after each environmental switch the diversifying BH quickly grew in frequency (highlighted by the increase of average standard deviation among the offspring of a single parent in  \textbf{Fig \ref{max_std}A}, and the second example phenotype in \textbf{Fig \ref{max_std}B}). However, in most cases this form of bet-hedging quickly got replaced by specialists and conservative BHs. 

We observed the initial success of the diversifying BH in all experiments where the season length was $\leq$ 100 generations, as well as at season length 300 with a medium or low mutation rate. At season length 500, this strategy was only observed in combination with a low mutation rate or low truncation size. Apart from a single run ($G = 500$, $m = 0.05$, $\mu$/pop size = 0.2), the diversifying strategy was lost eventually. In contrast, the incorporation of in-between values into the phenotype was observed in every experiment. However, as we increased the season length, mutation rate, or truncation size, less and less genes were half expressed after the 75 environmental switches, i.e., the conservative BH didn't appear or appeared later, while the diversifying BH reached higher frequencies and remained longer in the population. \textbf{In summary, season length of 300 was found to be ideal for the emergence and success of the diversifying BH, and increasing either the mutation rate or the truncation size favored the conservative strategy.}

Next, we looked at the \textbf{mutational robustness} of an evolved diversifying and conservative bet-hedger GRN, see \textbf{Fig \ref{mut_space}}. The conservative BH strategy was found to be considerably robust to random mutations. When the random mutations did change the phenotype, we found that it became even more of a conservative BH and less of a specialist. This robustness could have been the result of the sparse networks underlying the conservative BH phenotype (few edges with large weights, most edges 0 or small weight, data not shown).
On the other hand, the diversifying BH lost its ability to produce alternative phenotypes drastically. In most replicate experiments the mutated GRNs produced only one type of specialist after a few rounds of mutations. These networks were less sparse, though the degree distribution was similarly   power law.

\section{Discussion}

In this study, we investigated when and how the bet-hedger strategy evolves in a frequently changing environment. In contrast to previous work, we used an agent-based evolutionary algorithm of gene regulatory networks. This allowed us to evolve diversifying and conservative BHs without explicitly selecting for it, or having hard-coded strategies. We found that across different settings of the frequency of environmental change, strength of selection and mutation rates \textbf{diversifying BH evolved first, followed by conservative BH}. The above mentioned 3 parameters only changed the degree by which these strategies evolved, in a manner in line with results of previous studies. Higher frequency of environmental change favored the conservative \cite{Botero2015-ol, Mayer2017-lm}, smaller truncation size favored the diversifying BH strategy \cite{Tufto2015-lp, bull1987evolution}.

\begin{figure}
    \centering
    \includegraphics[width=\linewidth]{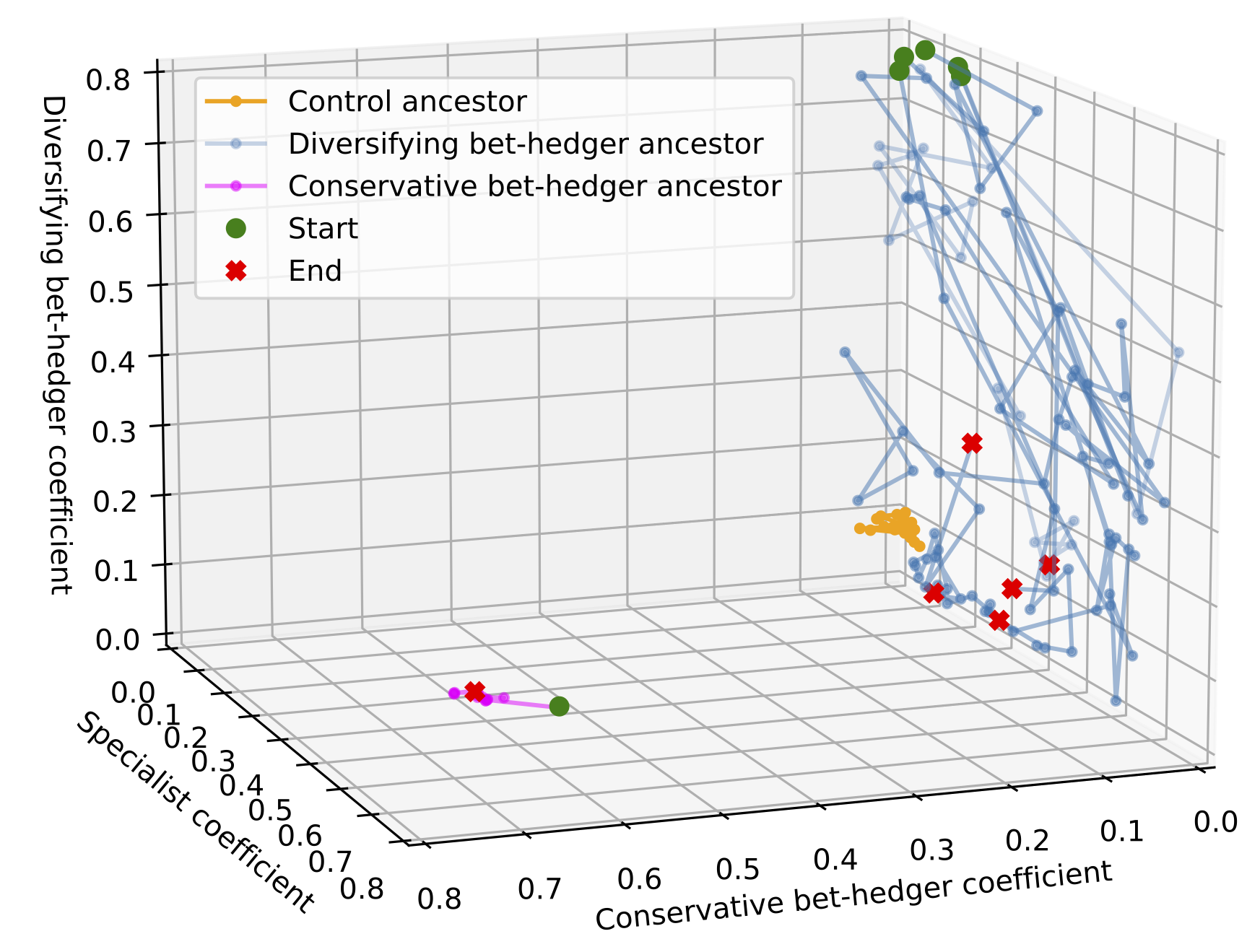}
    %\vspace*{-10mm}
    \caption{Effect of mutations on an evolved conservative and diversifying bet-hedger GRN in the 3D space of strategies. Lines represent the different random walks. The conservative BH is more robust than the diversifying BH.}
    \label{mut_space}
\end{figure}

At the beginning of every simulation, in the first environment, the populations quickly adapted and found the optimal solution. When the environment changed, suddenly the previously least fit individuals got to survive and reproduce while the previously fit lineages went extinct. After a huge drop in average fitness, the population adapted again to find the new fitness peak. The effect of a new mutation that causes the individual to be a bet-hedger, either by having an in-between phenotype or by having some proportion of the offspring be the opposite phenotype, is at first disadvantageous. If such mutation appears, it needs to stick around in the population despite being selected against until the environment changes. However, when the environment does change, the bet-hedger has a huge advantage over the specialist. The fitness of the conservative BH remains the same in-between value, which is now much higher than that of the specialist of the previous environment. Similarly, since the diversifying BH produces the alternative phenotype at some proportion, those individuals will now survive and reproduce. This explains why we saw the bet-hedging strategies increase in frequency right after the environment changed (\textbf{Fig \ref{max_std}}). While the conservative BH and specialist strategies are purely exploitative, diversifying BHs can be thought of an interesting solution to the tension between exploration and exploitation, in that there is structure to the variation that it creates. 

Our hypothesis for why we saw the initial success of the diversifying BH followed by its replacement by the conservative BH has to do with \textbf{how quickly the alternative strategies can be found and the mutational robustness of the evolved GRNs.} As the populations traverse the genotype space over the generations, going from parts of the landscape that produce one target phenotype to parts that produce the other phenotype, individuals can end up on the border of this high dimensional space where a few mutations push them into the other phenotype. Thus, the diversifying strategy could have been quickly found and selected for in our simulations for this reason, while the genotype that corresponds to an in-between phenotype could have been further away in the genotype space. Despite this advantage, our results suggest that the diversifying strategy is inherently more unstable. Offspring of a diversifying BH could easily become a specialist for the current environment due to random mutations, which then outcompetes the bet-hedger unless the environment switches right away. In contrast, once the conservative phenotype is found, it is robust to mutations, thus they are less likely to produce specialists that would drive them to extinction (\textbf{Fig \ref{mut_space}}). 

In conclusion, in response to adaptation to environmental variability, we observed the evolution of GRNs that were capable of generating the two alternative optimal phenotypes given random mutations (diversifying BHs), even without the implementation of gene duplication and deletion that was used in previous studies that found the evolution of this behavior in GRNs \cite{crombach2008evolution,Nichol2016-gh}. We also saw the evolution of the conservative BHs, as it outcompeted the diversifying strategy in most of our experiment. We argue that this dynamic could be explained by the robustness of the strategies.

%% The acknowledgments section is defined using the "acks" environment
%% (and NOT an unnumbered section). This ensures the proper
%% identification of the section in the article metadata, and the
%% consistent spelling of the heading.
\begin{acks}
This material is based upon work supported by the 2021-2022 University of Vermont Dr. Roberto Fabri Fialho Research Award to C.P. and the National Science Foundation Grant No. 2008413.
Computations were performed on the Vermont Advanced Computing Core supported in part by NSF Award No. 1827314.
\end{acks}

%%
%% The next two lines define the bibliography style to be used, and
%% the bibliography file.
\bibliographystyle{ACM-Reference-Format}
\bibliography{main}

%%
%% If your work has an appendix, this is the place to put it.
\appendix

\end{document}